\documentclass{article}
\usepackage{graphicx} 
\usepackage{natbib}
\usepackage{url}
\usepackage{float}
\usepackage{subcaption}
\usepackage{enumitem}
\usepackage{booktabs}
\usepackage{tabularx}
\usepackage{array}
\usepackage{hyperref}
\newcolumntype{Y}{>{\raggedright\arraybackslash}X}

\title{Nebulon Enterprise Simulated Threats for Phishing Research (NEST-Phish): \\ A Synthetic Enterprise Phishing Email Dataset for Behavioral and Machine-Learning Research}
\author{
Emily J. Winokur, Lauren S. Treiman, Allen G. Moore,\\
Paul Schutte, Joe Ingram, and Danielle N. Sanchez\thanks{Corresponding author: \href{mailto:dnsanc@sandia.gov}{dnsanc@sandia.gov}}
}
\date{September 2026}

\begin{document}

\maketitle

\begin{abstract}
Phishing remains one of the most persistent cyber threats, yet publicly shareable datasets for studying phishing in realistic enterprise email settings remain limited. To address this gap, we introduce a synthetic enterprise phishing email dataset built around a fictitious organization, Nebulon. The dataset spans a broad set of workplace communication themes and includes matched synthetic legitimate and phishing emails with interpretable phishing-cue annotations. Here, “legitimate” denotes the non-phishing class, not legitimately occurring organizational emails. Human-subject categorizations and classifier evaluations show that the dataset supports meaningful variation in phishing judgments while also providing learnable signal for supervised detection. This publicly released resource is intended to support future work on phishing detection, human susceptibility, explainability, and benchmark development in enterprise-like contexts.
\end{abstract}

\noindent\textbf{Keywords:} phishing, synthetic dataset, enterprise email, cybersecurity, human subjects, machine learning, email classification, behavioral research

\section{Introduction}
Phishing emails are deceptive messages that imitate legitimate communications in order to manipulate recipients into clicking malicious links, opening harmful attachments, disclosing sensitive information, or complying with fraudulent requests ~\citep{ftc_phishing, maheshwari2025phishing}. In 2024, phishing was one of the most persistent cyber threats and was the most frequently reported crime to the FBI Internet Crime Complaint Center with 193{,}407 complaints ~\citep{fbi_ic3_2024}.  This prevalence underscores its continued role as a primary vector for credential theft, fraud, and broader cyber compromise ~\citep{fbi_ic3_2024}. Phishing threats are also continually evolving, with emerging adversarial tactics including the use of large language models (LLMs) and QR-based attacks~\citep{weinz2025impact}. Technical defenses such as filtering and detection systems can reduce exposure, but they do not eliminate the threat entirely. Human judgment therefore remains an important line of defense, even though people are often treated as the “weakest link” in cybersecurity ~\citep{mitnick2003art,onwuegbuche2025securing}. For this reason, phishing research increasingly requires datasets that support both automated detection and the study of how people interpret, misclassify, and respond to suspicious emails~\citep{kavvadias2025understanding}.

Realistic enterprise email datasets are especially valuable for this purpose, but real organizational communications are difficult to share because of privacy concerns, operational sensitivity, and institutional restrictions~\citep{klimt2004enron}. Synthetic datasets therefore offer a practical alternative, provided they preserve the linguistic, thematic, and contextual properties needed for both human and machine evaluation ~\citep{lu2023machine}. However, publicly shareable datasets that combine enterprise-like themes, matched phishing and legitimate emails, controlled cue variation, and both human and machine evaluation remain limited. To address this gap, we introduce a synthetic enterprise phishing email dataset built around a fictitious organization, Nebulon. The dataset was designed to support both behavioral phishing research and machine-learning benchmarking. To support these goals, it incorporates realistic workplace themes, matched legitimate and phishing scenarios, controlled cue variation, and evaluation in both human-subject and classifier settings.

\section{Related Work and Existing Datasets}

Tables~\ref{tab:dataset_overview} and~\ref{tab:dataset_relevance} summarize some representative public datasets relevant to phishing and email classification research, including webpage- and URL-based phishing benchmarks~\citep{tan2018phishing,hannousse2021web}, classic email and spam corpora~\citep{klimt2004enron,sakkis2003memory,spamassassin2006}, phishing- and fraud-focused email collections~\citep{nazario2005nazario,tatman2002fraudulent,radev2008clair}, large merged email datasets~\citep{al2024novel,mendes2025meajor}, and more recent synthetic resources~\citep{kuladeep_p_2025,greco2024david}. Collectively, these datasets have supported important advances in phishing website detection, spam filtering, deceptive email analysis, and machine-learning-based email classification. Some are especially valuable because of their scale and realism, such as Enron~\citep{klimt2004enron} and MeAJOR~\citep{mendes2025meajor}, whereas others provide balanced benchmark tasks, structured feature representations, or synthetic phishing examples that are easier to share publicly~\citep{tan2018phishing,hannousse2021web,kuladeep_p_2025}. Although this review is not exhaustive, to the best of our knowledge no publicly shareable dataset currently combines enterprise-like themes, matched phishing and legitimate emails, controlled cue variation, and validation in both human and machine settings.

At the same time, the datasets summarized in Tables~\ref{tab:dataset_overview} and~\ref{tab:dataset_relevance} do not fully address the combination of needs motivating the present work. Website- and URL-focused datasets~\citep{tan2018phishing,hannousse2021web} do not capture email-level linguistic and contextual cues. Several email corpora contain only phishing emails or only legitimate emails, including Enron~\citep{klimt2004enron}, the Nazario Phishing Corpus~\citep{nazario2005nazario}, and the Fraudulent E-mail Corpus~\citep{tatman2002fraudulent,radev2008clair}, limiting their usefulness for matched phishing-versus-legitimate comparisons. Other datasets remain valuable for their intended purposes, but many are oriented toward related tasks such as spam filtering, fraud detection, or LLM authorship detection rather than matched enterprise phishing judgment ~\citep{sakkis2003memory,spamassassin2006,al2024novel,greco2024david}. Even when synthetic datasets are available~\citep{kuladeep_p_2025,greco2024david}, they may rely on repetitive templates or may not be designed to support realistic workplace themes, controlled phishing-cue variation, or human-subject evaluation. Together, these limitations point to the need for a publicly shareable synthetic enterprise email dataset designed to support both human and machine evaluation through realistic workplace themes, matched phishing and legitimate messages, and controlled phishing-cue variation.

\begin{table}[H]
\centering
\caption{Overview of representative public datasets relevant to phishing and email classification research. \(N\) denotes the number of stimuli in the dataset.}
\label{tab:dataset_overview}
\renewcommand{\arraystretch}{1.2}
\footnotesize
\setlength{\tabcolsep}{4pt}
\begin{tabularx}{\linewidth}{>{\raggedright\arraybackslash}X p{1.7cm} p{1.0cm} >{\raggedright\arraybackslash}X}
\hline
\textbf{Dataset} & \textbf{Type} & \textbf{\(N\)} & \textbf{Coverage} \\
\hline
Tan (2018) \citep{tan2018phishing} & Web features & 10,000 & Phishing + legitimate \\
Hannousse and Yahiouche (2021) \citep{hannousse2021web} & URL features & 11,430 & Phishing + legitimate \\
Enron \citep{klimt2004enron} & Email & 200,399 & Legitimate only \\
Ling-Spam \citep{sakkis2003memory} & Email & 2,893 & Spam + legitimate \\
SpamAssassin \citep{spamassassin2006} & Email & 4,000+ & Spam + legitimate \\
Fraudulent E-mail Corpus \citep{tatman2002fraudulent,radev2008clair} & Email & 2,500+ & Fraud/scam only \\
Nazario Phishing Corpus \citep{nazario2005nazario} & Email & 11,527 & Phishing only \\
Al-Subaiey et al. (2024) \citep{al2024novel} & Email & 80,000+ & Phishing/spam + legitimate \\
MeAJOR \citep{mendes2025meajor} & Email & 135,894 & Phishing + legitimate \\
Kuladeep (2025) \citep{kuladeep_p_2025} & Email & 10,000 & Phishing + legitimate \\
Greco et al. (2024) \citep{greco2024david} & Email & --- & Human- and LLM-generated phishing \\
\hline
\end{tabularx}
\end{table}

\begin{table}[H]
\centering
\caption{Summary of the relevance of representative public datasets to the present work.}
\label{tab:dataset_relevance}
\renewcommand{\arraystretch}{1.2}
\footnotesize
\setlength{\tabcolsep}{5pt}
\makebox[\textwidth][c]{%
\begin{tabular}{p{4.6cm} p{4.0cm} p{5.0cm}}
\hline
\textbf{Dataset} & \textbf{Strength} & \textbf{Main limitation for this work} \\
\hline
Tan (2018) \citep{tan2018phishing} & Balanced benchmark for phishing website classification & Focuses on website-level features rather than email content, enterprise scenarios, or user judgments \\
Hannousse and Yahiouche (2021) \citep{hannousse2021web} & Useful benchmark for machine-learning-based phishing website detection & Not designed for full email content, workplace phishing scenarios, or contextual human judgments \\
Enron \citep{klimt2004enron} & Foundational corpus of real organizational email & Contains no phishing emails and is not suitable as a directly shareable phishing stimulus set \\
Ling-Spam \citep{sakkis2003memory} & Widely used benchmark for spam filtering & Mailing-list domain and older communication style reduce relevance to modern enterprise phishing \\
SpamAssassin \citep{spamassassin2006} & Public benchmark with messages of varying difficulty & Designed for general spam filtering rather than enterprise phishing or controlled cue studies \\
Fraudulent E-mail Corpus \citep{tatman2002fraudulent,radev2008clair} & Useful source of deceptive email text and metadata & Focuses on advance-fee fraud rather than contemporary enterprise phishing \\
Nazario Phishing Corpus \citep{nazario2005nazario} & Widely used public source of phishing examples & Lacks legitimate counterparts and enterprise context \\
Al-Subaiey et al. (2024) \citep{al2024novel} & Large corpus with message text and metadata & Heterogeneous formatting and noisy text make controlled cue analysis difficult \\
MeAJOR \citep{mendes2025meajor} & Large, clearly structured feature-rich phishing email corpus & Not designed for matched scenarios, controlled cues, or human-subject realism \\
Kuladeep (2025) \citep{kuladeep_p_2025} & Structured synthetic corpus with phishing type and severity labels & Within-theme messages appear templated, limiting subtle cue variation and realism \\
Greco et al. (2024) \citep{greco2024david} & Useful for detecting AI-generated phishing content & Oriented toward authorship detection rather than balanced phishing-versus-legitimate classification \\
\hline
\end{tabular}%
}
\end{table}

\section{Present Work and Contributions}

Our synthetic enterprise phishing email dataset is designed to support both behavioral phishing research and machine-learning benchmarking. Emails were generated using a previously developed synthetic dataset, subject matter expert input, and LLM-assisted generation, then organized into workplace-relevant themes such as information technology, finance, compliance, document sharing, shipping, travel, training, physical access, and corporate communications. Legitimate and phishing emails were constructed as matched pairs with similar scenarios, tone, and subject matter, while systematically varying phishing-related cues. By combining matched legitimate/phishing scenarios with systematic variation in phishing-related cues, the dataset was intended to include both relatively obvious and more ambiguous examples, rather than only clearly phishing and clearly legitimate messages.

The dataset was also subjected to iterative human review prior to participant assessment. The LLM-generated emails were reviewed by multiple team members for plausibility, duplication, ambiguity, and cue coherence, and problematic items were revised or removed before the final participant-facing set was established. We then evaluated the dataset in an online human-subject study in which participants completed a brief phishing training module and subsequently classified emails as phishing or legitimate while also providing confidence ratings. These judgments allowed us to assess both overall accuracy and performance as a function of email type, phishing-related signal count, workplace theme, and subjective certainty, helping us evaluate whether the dataset captured meaningful variation in classification difficulty. Because the dataset was also intended to support machine-learning benchmarking, we additionally evaluated supervised classifiers on the dataset to assess whether it contains learnable structure for phishing detection.

Overall, this work introduces a publicly shareable synthetic enterprise phishing email dataset that addresses limitations of prior corpora by combining realistic workplace communication themes, closely matched legitimate and phishing emails, controlled cue variation, and validation in both human-subject and machine-learning settings. The goal is not to replace real organizational email corpora, but rather to provide a research-useful resource that is shareable, controlled, and behaviorally informative. By combining interpretable phishing cues, matched legitimate/phishing scenarios, human evaluation, and baseline machine-learning benchmarks, the dataset is intended to support future work on phishing detection, human susceptibility, explainability, and benchmark development.

\section{Dataset Construction \& Validation}
\subsection{Seed Data and Design Goals}

The dataset was constructed using three primary inputs: (1) a synthetic email dataset developed by our group for a previous phishing-related experiment, (2) subject matter expert (SME) input, and (3) LLM-assisted generation with ChatGPT 5.2. The earlier dataset had been developed internally for a prior phishing experiment and consisted of 40 synthetic phishing and legitimate email stimuli. The legitimate stimuli were derived in part from messages in the Enron Email Dataset ~\citep{klimt2004enron}, and the phishing stimuli were developed using subject matter expertise and public phishing examples, with structured cue annotations that made the dataset a useful scaffold for the present work Rather than generating emails without constraints, we used this earlier synthetic dataset as a structured starting point in order to preserve realistic enterprise-style formats, scenarios, and message components while extending the thematic breadth and difficulty of the email set.

Here, the seed dataset refers to the earlier internally developed dataset used as the structured starting point for the present corpus. It contained both message content and structured annotations. Each seed email included metadata fields such as date, sender, recipient, subject line, and body text, along with a reference label indicating whether the email was phishing or legitimate. In addition, the seed set included binary phishing-cue indicators for several salient features: the presence of a link, the presence of an attachment, a generic greeting, urgent language, an external sender, and a request for personal information. These fields provided a scaffold for generation and later analysis by anchoring new emails to an explicit set of interpretable message characteristics. The seed dataset also served as a reference point for maintaining plausible cue distributions in the expanded corpus.

There were several design goals in constructing the new dataset. First, the emails were intended to approximate realistic workplace communications rather than being dominated by consumer-phishing examples such as package-delivery scams, prize notifications, or generic account-warning emails tied to consumer services. Second, the dataset was designed to include controlled variability in both content and phishing-cue structure, allowing the resulting emails to span a range of classification difficulty. Third, legitimate and phishing emails were constructed in a roughly balanced way, often as matched pairs within the same general theme or scenario, so that the dataset would support both human-subject studies and machine-learning benchmarks. Finally, because the dataset was intended for public release, all emails were synthetic and designed to avoid disclosure of real organizational communications or sensitive operational information. Accordingly, the "legitimate" emails in the dataset should be understood as non-phishing enterprise-style messages, rather than organizational emails that occurred naturally.

\subsection{Themes}

The dataset was designed to cover a broad range of enterprise-relevant communication contexts rather than a narrow set of generic phishing scenarios. The theme set was iteratively refined based on themes represented in the seed dataset and in corporate phishing detection exercises, with the goal of covering a broad range of enterprise-relevant email contexts. Themes were refined over multiple rounds of generation and review to broaden realism and diversify the contexts represented in the final corpus. To support both behavioral studies and benchmarking tasks, emails were organized into themes that reflect common categories of workplace communication, including information technology, finance, compliance, operations, human resources, document sharing, and organizational announcements. These themes were intended to capture both routine benign communications and plausible phishing attempts embedded within realistic business workflows.

The initial theme set focused on core enterprise scenarios such as account and security notifications, information technology (IT) maintenance, human resources (HR) and payroll administration, compliance acknowledgments, finance and vendor billing, operations and dispatch communication, document sharing prompts, shipping notifications, and voicemail alerts. Additional themes were later added to broaden realism and diversify the range of contexts represented in the dataset, including travel logistics, physical access and security, employee giving, employee wellness, survey or prize promotion, training and certifications, data privacy, and corporate communications. The full set of themes is listed in Table~\ref{tab:theme_list}.

\begin{table}[H]
\centering
\caption{Enterprise communication themes represented in the dataset. Acronyms are expanded for clarity.}
\label{tab:theme_list}
\renewcommand{\arraystretch}{1.2}
\begin{tabular}{l}
\hline
\textbf{Theme} \\
\hline
Account/security notifications (information technology/authentication) \\
Information technology maintenance / access continuity \\
Human resources/payroll/benefits administration \\
Compliance/policy acknowledgements \& surveys \\
Finance/accounts payable/vendor billing \\
Operations/field/grid dispatch \\
Document sharing/collaboration prompts \\
Shipping/delivery notifications \\
Voicemail/communications system alerts \\
Travel/calendar logistics \\
Physical access/security \\
Employee giving / donations campaign \\
Employee wellness challenge \\
Survey / prize promotion \\
Training \& certifications \\
Data privacy / records management \\
Corporate communications / executive announcements \\
\hline
\end{tabular}
\end{table}

\subsection{Phishing Signal Taxonomy}

To support both controlled dataset construction and downstream analysis, we defined a taxonomy of phishing-related cues that combined a small set of explicit binary indicators with a broader set of realistic textual and structural artifacts. This taxonomy was informed in part by the National Institute of Standards and Technology (NIST) Phish Scale~\cite{dawkins2023phishing,steves2020categorizing}, but it does not constitute a full implementation of that framework. Several of our indicators were chosen to align with NIST cue categories, including attachment-related cues, generic greeting, urgent language, and requests for sensitive information, while others, such as link presence and apparent external-sender status, were included as simplified operational features for use in a synthetic enterprise email setting. Some NIST components were not included as structured fields, such as visual presentation, branding cues and the premise-alignment component.

The core binary indicators were selected because they are both interpretable and commonly discussed in phishing awareness training: whether the email contained a link, whether it included an attachment, whether it used a generic greeting, whether it contained urgent language, whether it appeared to come from an external sender, and whether it requested personal information ~\citep{dawkins2023nistphishscale, ftc_phishing, steves2020categorizing}. These indicators were retained as structured fields in the dataset and served as the primary cue variables for generation, descriptive statistics, and later analysis of participant and classifier behavior.

These binary indicators were included as interpretable cues, not as deterministic markers of phishing. In realistic workplace communication, links, attachments, generic greetings, and urgent language can all appear in legitimate messages as well as phishing emails. For this reason, the dataset was designed so that these features contributed to overall cue structure without functioning as simple classification rules. For example, the \texttt{has\_request\_personal\_information} field was used to mark emails that explicitly requested personal or sensitive information, especially when such requests were inappropriate to the email context or delivery channel.

In addition to the core binary fields, we incorporated a broader set of realistic phishing cues that were reflected in the email text, sender information, links, or attachment names ~\citep{dawkins2023nistphishscale, ftc_phishing, steves2020categorizing}. These included use of \texttt{http} rather than \texttt{https}, suspicious or macro-enabled attachment formats (e.g., \texttt{.docm}) or deceptive attachment names with double extensions (e.g., \texttt{.pdf.exe} or \texttt{.pdf.zip}), deceptive or lookalike domains, mismatches between the sender and other details in the email body, sender-role mismatches (e.g., an IT sender requesting payroll-related information), threatening or coercive language, and unusual or inappropriate workflow requests. These additional cues were used to increase realism and vary difficulty, but in order to preserve a compact and interpretable annotation schema, they were not encoded as separate dataset columns.

This taxonomy also informed the participant training module used in the human evaluation study (as described in Section \ref{sec:training}). Participants were trained on a subset of these cues before completing the classification task (Section \ref{fig:training_module}). As a result, the taxonomy served both as a generation framework and a conceptual bridge between dataset design, participant instruction, and later cue-based analyses of human and model performance.

\subsection{LLM-Assisted Email Generation}

After defining the structure of the seed dataset, the enterprise themes, and the phishing signal taxonomy, we used ChatGPT 5.2 to generate new emails in a controlled, human-in-the-loop manner. The seed dataset was provided to the model in structured text form, including the sender, recipient, date, subject, and body for each email, along with binary indicators specifying whether the email contained a link, an attachment, a generic greeting, a request for personal information, urgent language, or an external sender. These examples served as templates for plausible workplace tone, level of detail, message structure, and phishing-cue combinations, rather than as content to be copied directly. In this workflow, the model generated one legitimate/phishing email pair at a time. We then reviewed, revised, and redirected generation after each step before requesting the next pair. Within each pair, the legitimate and phishing emails shared the same general theme and scenario and were written to be similar in subject matter, tone, and contextual framing. For example, a legitimate and phishing email might both concern a payroll adjustment, a shared document, a travel update, or a system maintenance notification. As each pair was generated, it was manually reviewed and, when necessary, revised to ensure that the legitimate and phishing versions were comparable in overall scenario while still differing in meaningful phishing-related ways. This paired construction helped reduce topic-level confounds and made it possible to compare human and classifier performance on emails that were similar in content but differed in cue structure.

LLM prompting also emphasized controlled variation in phishing difficulty. Not all phishing emails were designed to contain the same signals, and not all legitimate emails were intentionally cue-free. Instead, we iteratively adjusted prompting and email content so that some phishing emails contained multiple overt indicators (e.g., an external sender, urgent language, and a suspicious attachment), whereas others relied on subtler or mixed signals (e.g., an internal-looking sender paired with an inappropriate request for personal information, or a sender/link mismatch without explicit threat language). Likewise, some legitimate emails were allowed to contain superficially suspicious features, such as links, generic greetings, external senders, or deadline language, in order to better reflect realistic workplace communication and avoid making classification trivially easy. When generated emails appeared unrealistically simple, overly repetitive, or insufficiently distinct from their paired counterpart, they were manually edited, partially rewritten, or regenerated.

The generation process was therefore iterative rather than one-shot. Additional prompts were used throughout construction to broaden theme coverage, improve realism, reduce repetition, and better align cue prevalence with the seed corpus. For example, prompts were revised when too many phishing emails relied on the same cue combination, when a phishing email seemed insufficiently distinguishable from its legitimate counterpart, or when a legitimate email appeared unrealistically devoid of common workplace features such as attachments or external senders. Later rounds of prompting and editing also introduced subtler artifacts, such as \texttt{http} links, suspicious attachment naming conventions, sender-role mismatches, lookalike domains, and threatening or coercive language. In this way, the LLM functioned as a structured generation assistant under continuous human supervision rather than as an autonomous source of finalized dataset content.

\subsection{Internal Review and Quality Control}

As an additional layer of validation, and prior to use in the participant study or classifier analyses, all emails underwent internal review by multiple team members. Using a simple interface that displayed emails in a realistic format, reviewers examined each email individually, categorized it as phishing or legitimate, rated their confidence, and provided qualitative notes. These notes were used to flag examples that were ambiguous, unrealistic, overly repetitive, too similar to other emails, insufficiently differentiated from their legitimate/phishing counterpart, or inconsistent with their assigned cue indicators. This review process helped identify duplicate or near-duplicate items, refine the intended range of difficulty, and preserve thematic diversity without over-representing highly similar scenarios or cue combinations.

Emails flagged during internal review were manually examined and either revised or removed from the dataset. Depending on the nature of the issue, revisions ranged from minor wording edits and cue-label corrections to more substantial rewriting or replacement of the email. This quality-control process was iterative, with reviewers examining updated versions of the dataset across multiple rounds of refinement to improve realism, internal consistency, and difficulty variability. Because this review was conducted internally, a larger participant sample was still needed to validate how the emails functioned as classification stimuli and to assess whether the intended variation in difficulty was reflected in human judgments.

\subsection{Initial Downselection of Candidate Emails}
\label{sec:downselection}
Before analyzing participant performance, we conducted an initial manual review of the generated candidate email set to remove items whose intended signal structure was not sufficiently interpretable. This review was designed to identify emails in which the phishing or legitimate label was undermined by artifact-driven ambiguity rather than realistic task difficulty. In practice, excluded items tended to fall into a small number of categories: (1) overly ambiguous or internally contradictory cue combinations, (2) sender--link combinations that appeared legitimately plausible or otherwise conflicted with the intended label, (3) misspellings or domain-formatting artifacts that introduced unintended interpretations, and (4) unclear internal versus external organizational context within the fictitious Nebulon setting.

This downselection step was intended to preserve challenging but coherent examples while removing emails whose ambiguity arose primarily from construction artifacts. Of the 169 candidate emails initially reviewed, 23 were excluded during this process, leaving 146 emails in the refined set used for all subsequent participant and classifier analyses. The signal presence across the downselected legitimate and phishing emails is displayed in Table \ref{tab:cue_distribution_final}, and the final theme distribution is displayed in Table \ref{tab:theme_distribution_final}.

\

\subsection{Final Dataset Characteristics}

The final retained dataset consisted of 146 synthetic emails used for all subsequent participant and classifier analyses. Each email record included message metadata (e.g., date, sender, recipient, subject, and body), a ground-truth label, a theme label, binary phishing-cue indicators, and annotation notes used during dataset construction and review. The retained dataset remained approximately balanced by class, with 77 legitimate emails and 69 phishing emails distributed across 17 enterprise-related themes (Table ~\ref{tab:theme_distribution_final}). The distribution of annotated phishing-cue indicators in the retained dataset is shown in Table ~\ref{tab:cue_distribution_final}. Theme counts varied across categories, with larger representation in finance/accounts payable/vendor billing, account/security notifications, human resources/payroll/benefits administration, and operations/field/grid dispatch, and smaller representation in corporate communications, voicemail alerts, and physical access/security. This distribution was intended to provide broad thematic coverage while preserving a manageable and interpretable set of analysis stimuli.

\begin{table}[H]
\centering
\caption{Theme distribution in the final retained dataset after manual downselection.}
\label{tab:theme_distribution_final}
\renewcommand{\arraystretch}{1.2}
\resizebox{\textwidth}{!}{%
\begin{tabular}{lccc}
\hline
\textbf{Theme} & \textbf{Legit} & \textbf{Phish} & \textbf{Total} \\
\hline
Finance/accounts payable/vendor billing & 8 & 8 & 16 \\
Account/security notifications (information technology/authentication) & 9 & 5 & 14 \\
Human resources/payroll/benefits administration & 6 & 6 & 12 \\
Operations/field/grid dispatch & 7 & 5 & 12 \\
Compliance/policy acknowledgements \& surveys & 5 & 5 & 10 \\
Shipping/delivery notifications & 5 & 5 & 10 \\
Travel/calendar logistics & 6 & 4 & 10 \\
Document sharing/collaboration prompts & 6 & 3 & 9 \\
Information technology maintenance / access continuity & 4 & 5 & 9 \\
Employee wellness challenge & 4 & 3 & 7 \\
Survey / prize promotion & 3 & 4 & 7 \\
Employee giving / donations campaign & 3 & 3 & 6 \\
Training \& certifications & 3 & 3 & 6 \\
Data privacy / records management & 2 & 3 & 5 \\
Physical access/security & 3 & 2 & 5 \\
Corporate communications / executive announcements & 2 & 2 & 4 \\
Voicemail/communications system alerts & 1 & 3 & 4 \\
\hline
\textbf{Total} & \textbf{77} & \textbf{69} & \textbf{146} \\
\hline
\end{tabular}%
}
\end{table}

The email metadata included six binary phishing-cue indicators that were used both in construction and later analysis: links, attachments, generic greetings, urgent language, external senders, and requests for personal information (Table ~\ref{tab:cue_distribution_final}). As expected, several phishing-associated cues appeared more frequently in phishing than in legitimate emails, including urgent language, external senders, and requests for personal information. Other cues, such as links, attachments, and generic greetings, appeared in both classes, allowing the dataset to include both more obvious phishing emails and more ambiguous legitimate messages that shared phishing-like surface features. 


\begin{table}[H]
\centering
\caption{Distribution of binary phishing-cue indicators in the final retained dataset. Values are counts with within-label percentages.}
\label{tab:cue_distribution_final}
\renewcommand{\arraystretch}{1.2}
\begin{tabular}{lcc}
\hline
\textbf{Cue} & \textbf{Legit} & \textbf{Phish} \\
\hline
Link present & 40 (51.9\%) & 46 (66.7\%) \\
Attachment present & 19 (24.7\%) & 19 (27.5\%) \\
Generic greeting & 35 (45.5\%) & 42 (60.9\%) \\
Urgent language & 8 (10.4\%) & 43 (62.3\%) \\
External sender & 14 (18.2\%) & 51 (73.9\%) \\
Request for personal information & 5 (6.5\%) & 22 (31.9\%) \\
\hline
\end{tabular}
\end{table}

\section{Human Evaluation}
\subsection{Participant Recruitment and Characteristics}

Thirty participants (20 female, 10 male) were recruited from Prolific \citep{palan2018prolific}, an online research recruitment platform (\url{www.prolific.com}). All participants resided in the United States, were at least 18 years old (\(M = 35.1\), \(SD = 11.2\)), fluent in English, had a Prolific approval rating greater than 90\%, and had completed at least 10 previous Prolific submissions. The mean participation time was 65.5 minutes, and participants were paid \$12 for their time. All procedures were approved by the Human Subjects Board at Sandia National Laboratories (Protocol No. SNL000571), and informed consent was obtained from all participants.

\subsection{Participant Training}
\label{sec:training}

Before beginning the email categorization task, participants completed a brief interactive training module on common characteristics of phishing emails. The training introduced nine phishing indicators: (1) generic greetings, (2) lack of personalization, (3) urgent or threatening language, (4) strange or misspelled URLs, (5) mismatches between the sender and other email details (e.g., links, signatures, or referenced organizations), (6) use of HTTP instead of HTTPS, (7) suspicious attachments (particularly double extensions such as \texttt{.pdf} or \texttt{.zip} or executable formats such as \texttt{.exe} or \texttt{.msi}), (8) requests for personal information (e.g., Social Security number, bank account information, or date of birth), and (9) grammatical errors. Although grammatical errors were included as a commonly discussed phishing cue in the training module, they were not strongly represented in the present dataset because the emails were designed to reflect more plausible enterprise-style phishing, where obvious writing errors are not always present and can make messages easier to detect.

The module presented these indicators in an interactive format (Figure~\ref{fig:training_module_a}). Participants clicked through each tip individually to read a short explanation and view an example email adapted from prior phishing examples \citep{norton_phishing_examples} illustrating how the indicator could appear in practice (Figure~\ref{fig:training_module_b}). Participants were required to review all nine indicators before proceeding. To reinforce the material, they then completed a seven-question quiz covering the training content (Appendix~\ref{apx:trainingquiz}). Participants were not required to answer all questions correctly; however, they received immediate feedback on each response, telling them the correct answer and providing a short explanation.

\subsection{Procedure}

\begin{figure}[H]
    \centering
\includegraphics[width=\textwidth,height=0.8\textheight,keepaspectratio]{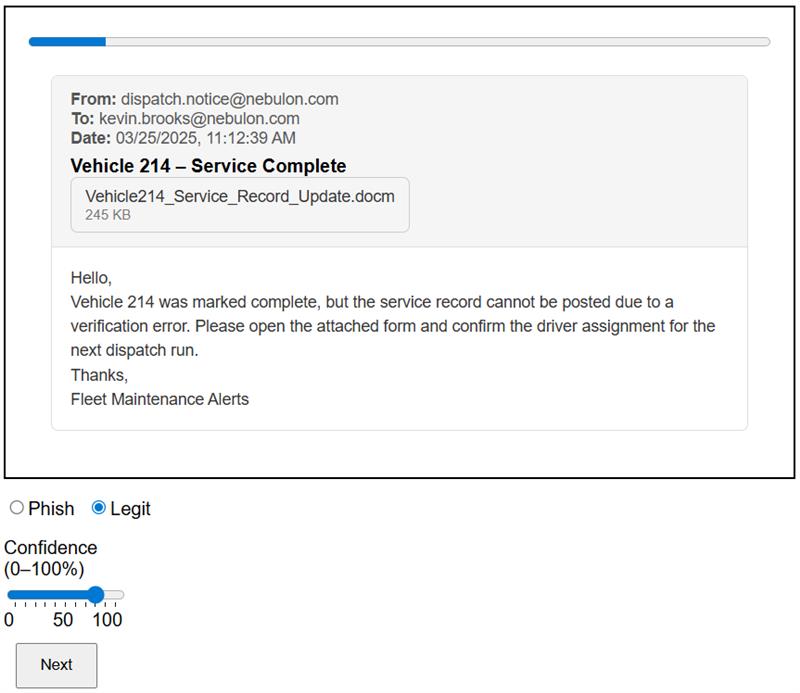}
    \caption{Example email interface shown to participants. For each email, participants indicated whether the email was legitimate or phishing and their level of confidence in the categorization.}
    \label{fig:user_interface_example}
\end{figure}

Each participant viewed and categorized 172 emails (85 legitimate, 84 phishing, and 3 attention check emails). All participants answered the three attention checks correctly. For each email, participants categorized the email as phishing or legitimate then indicated their confidence in the categorization on a scale of 0--100 in 10-point increments, where a higher number represented greater confidence. An example email is shown in Figure ~\ref{fig:user_interface_example}. For the attention check emails, participants were told how to categorize the email and the level of confidence to report. All emails were presented in a random order, where the order differed for every participant, but each participant saw every email. Because each participant evaluated the full retained email set, the study yielded 4,380 trial-level judgments that supported item-level analyses of accuracy, confidence, signal count, and theme. Participants did not receive feedback on their accuracy at any point during the task. 

\subsection{Overall Participant Performance on the Refined Set}

In total, the analysis included 30 participants with 146 emails per participant, resulting in a total of 4,380 judgments. Overall human categorization accuracy was 82.1\%, indicating that participants performed above chance level on the email categorization task but were not perfect. Performance differed by true email label. Participants correctly classified legitimate emails with 88.8\% accuracy, compared with 74.7\% accuracy for phishing emails. This means that 25.3\% of phishing-emails were labeled as legitimate, whereas 11.2\% of legitimate-emails were labeled as phishing (Table~\ref{tab:overall_performance}). This pattern suggests that participants were more likely to overlook phishing emails than to incorrectly flag legitimate emails.

\begin{table}[H]
\centering
\caption{Overall participant performance on the refined email set.}
\label{tab:overall_performance}
\begin{tabular}{lr}
\hline
\textbf{Metric} & \textbf{Value} \\
\hline
Overall accuracy & 82.1\% \\
Accuracy on phishing emails & 74.7\% \\
Accuracy on legitimate emails & 88.8\% \\
Phishing emails labeled as legitimate & 25.3\% \\
Legitimate emails labeled as phishing & 11.2\% \\
\hline
\end{tabular}
\end{table}

\begin{figure}[t]
    \centering
    \includegraphics[width=0.8\linewidth]{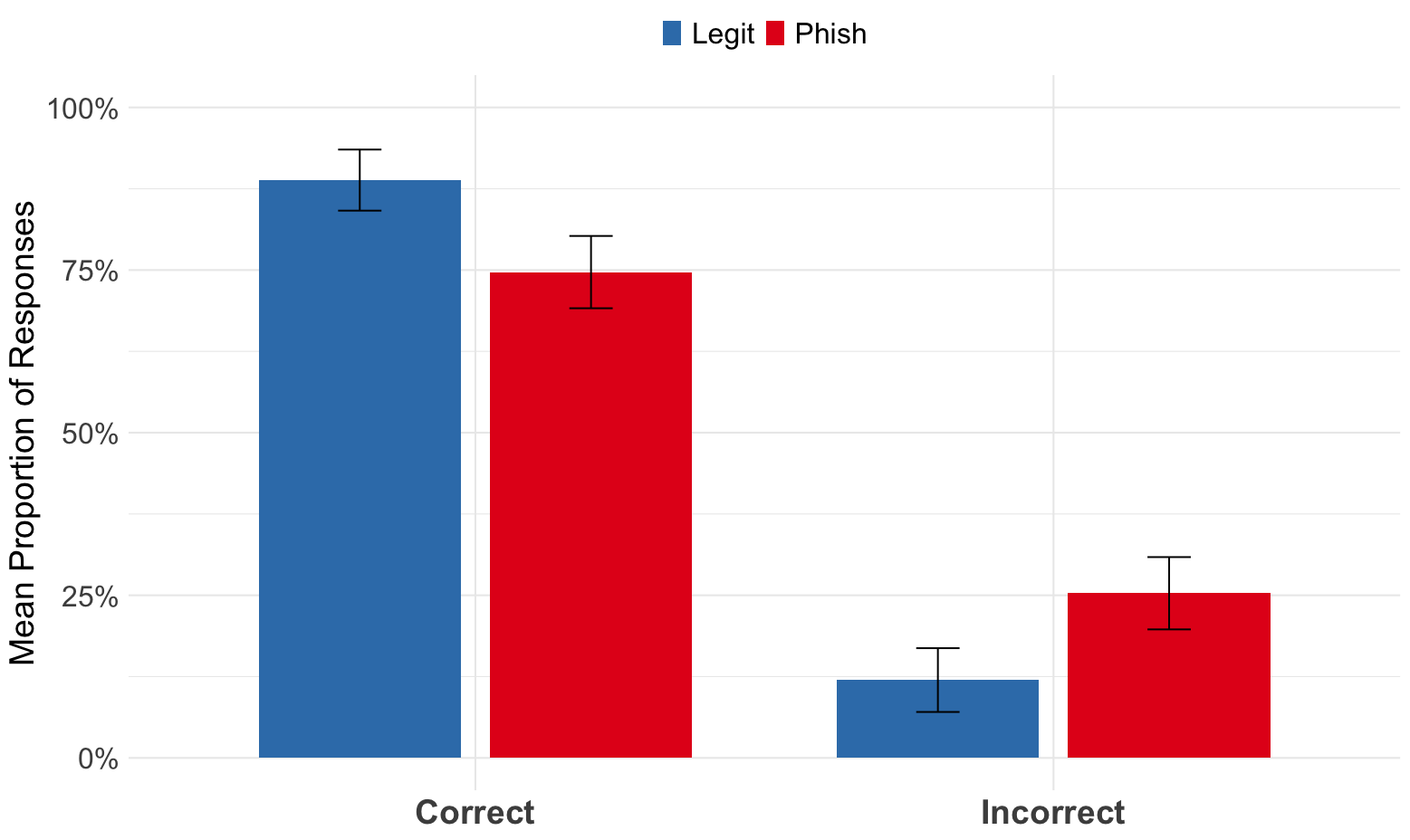}
    \caption{Mean proportion of correct and incorrect responses for phishing and legitimate emails. Error bars represent 95\% confidence intervals across participants. For phishing emails, incorrect responses indicate that the email was labeled as legitimate; for legitimate emails, incorrect responses indicate that the email was labeled as phishing.}
    \label{fig:accuracy_by_type}
\end{figure}

To test whether accuracy differed by email type, we fit a mixed-effects logistic regression predicting trial-level accuracy from true email label, with random intercepts for participant and email. The model confirmed that email type significantly affected performance, \(\chi^2(1)=37.52\), \(p<.001\). Specifically, the odds of a correct response were lower for phishing emails than for legitimate emails (\(OR=0.35\), 95\% CI [0.26, 0.48]; Figure~\ref{fig:accuracy_by_type}), indicating that the odds of a correct response were approximately 65\% lower for phishing emails than for legitimate emails.

\subsection{Performance Varies with Phishing-Related Signal Count}

\begin{figure}[t]
    \centering
    \includegraphics[width=0.85\linewidth]{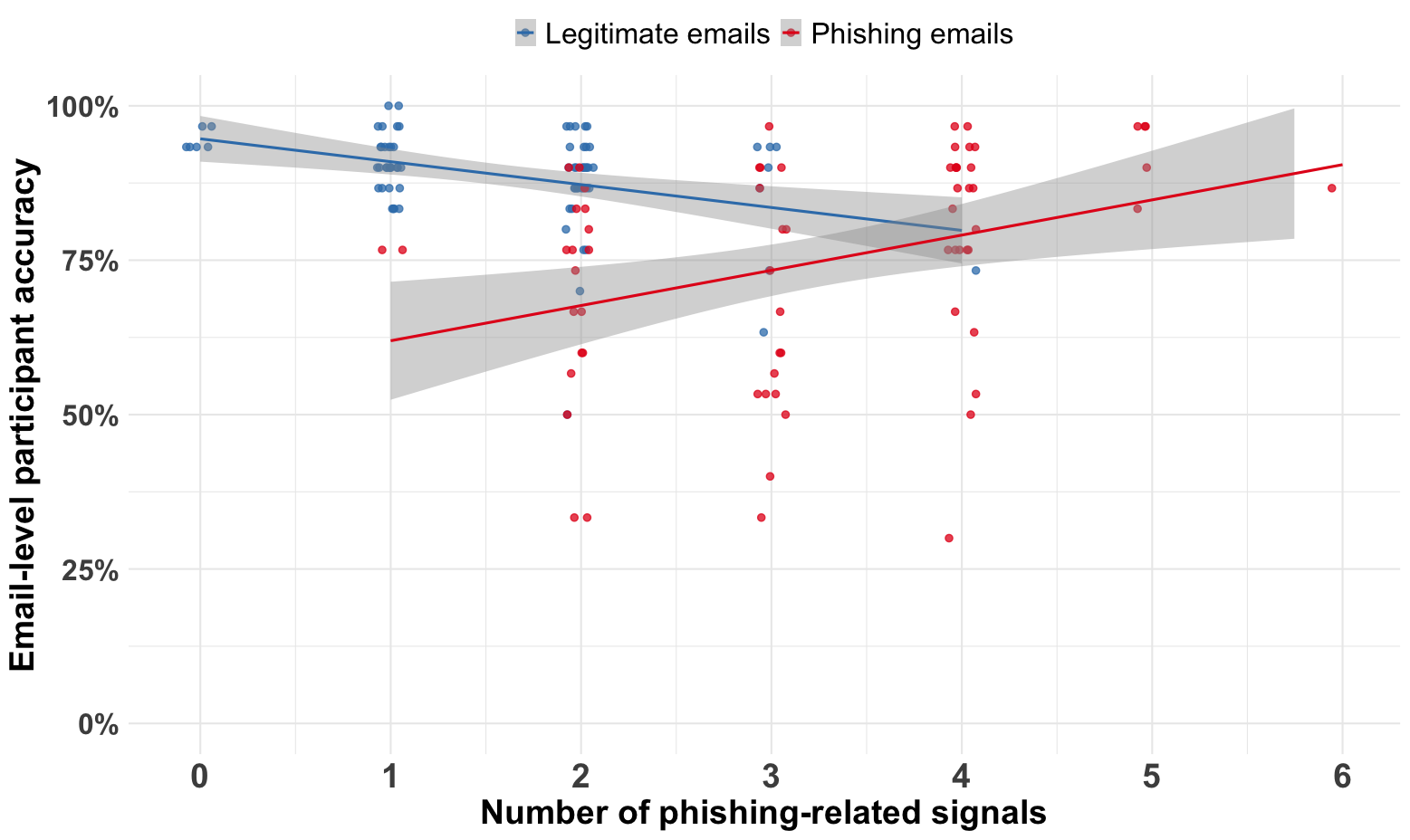}
    \caption{Email-level participant accuracy as a function of phishing-related signal count, shown separately for phishing and legitimate emails. Each point represents one email, with accuracy calculated as the proportion of participants who classified that email correctly. Trend lines illustrate the relationship between signal count and classification accuracy within each email type.}
    \label{fig:signal_count_accuracy}
\end{figure}

To examine whether cue density influenced performance, we defined a signal-count measure as the sum of six annotated phishing-related features: links, attachments, generic greetings, urgent language, external senders, and requests for personal information. As shown in Table~\ref{tab:supp_signal_count_distribution}, legitimate emails were concentrated at lower signal counts, whereas phishing emails were concentrated at higher signal counts. To test whether the association between signal count and accuracy differed by email type, we fit a mixed-effects logistic regression predicting trial-level accuracy from signal count, true email label, and their interaction, with random intercepts for participant and email. The model showed a significant interaction between signal count and true email label, \(\chi^2(1)=20.09\), \(p<.001\), indicating that the effect of signal count on participant accuracy differed for phishing and legitimate emails (Figure~\ref{fig:signal_count_accuracy}).

Specifically, for legitimate emails, each additional phishing-related signal was associated with lower odds of a correct response (\(\beta=-0.38\), \(SE=0.13\), \(OR=0.68\), \(p=.004\)), corresponding to an approximately 31.6\% decrease in the odds of correct classification per additional signal. For phishing emails, the interaction term was positive (\(\beta=0.76\), \(SE=0.17\), \(OR=2.14\), \(p<.001\)), indicating that additional signals made phishing emails easier to classify correctly rather than harder (Figure ~\ref{fig:signal_count_accuracy}). This pattern suggests that the dataset contains both relatively obvious phishing emails and more ambiguous legitimate emails that share phishing-like features, making it useful for studying both successful detection and miscategorization of legitimate emails. 

\subsection{Performance Across the Enterprise Themes}

The refined dataset spans 17 enterprise-related themes representing realistic emails that an employee of an energy company might receive (Table~\ref{tab:theme_accuracy_by_label}). To test whether accuracy differed across themes after accounting for email type, we fit a mixed-effects logistic regression predicting trial-level accuracy from true email label and theme, with random intercepts for participant and email. The model did not identify a significant overall main effect of theme, \(\chi^2(16)=14.60\), \(p=.554\). This suggests that theme-related variation in accuracy was not well captured by a single omnibus effect once the overall difference between phishing and legitimate emails was taken into account.

However, when phishing and legitimate emails were examined separately (Table~\ref{tab:theme_accuracy_by_label}; see also Appendix Figure~\ref{fig:theme_accuracy_by_label_supp}), legitimate emails were classified more accurately than phishing emails in 16 of the 17 themes. The largest observed differences between legitimate and phishing accuracy were in the physical access/security, data privacy/records management, and training/certifications themes, where legitimate emails were generally identified reliably but phishing emails were substantially harder to detect. Survey / prize promotion was the main exception: phishing emails in this theme were classified more accurately than legitimate emails, suggesting that legitimate messages in this category may have shared characteristics commonly associated with suspicious promotional content. Taken together, these descriptive patterns suggest that susceptibility to phishing may differ across workplace contexts, making the dataset useful for studying context-specific phishing vulnerability even though theme did not emerge as a significant overall predictor of accuracy in the mixed-effects model.

\begin{table}[H]
\centering
\caption{Theme-level participant accuracy by true email label. Percent values are mean participant accuracies within each theme.}
\label{tab:theme_accuracy_by_label}
\renewcommand{\arraystretch}{1.2}
\resizebox{\textwidth}{!}{%
\begin{tabular}{lcccc}
\hline
\textbf{Theme} & \textbf{Legit \(n\)} & \textbf{Legit \%} & \textbf{Phish \(n\)} & \textbf{Phish \%} \\
\hline
Physical access/security & 3 & 86.7\% & 2 & 41.7\% \\
Data privacy / records management & 2 & 86.7\% & 3 & 61.1\% \\
Training \& certifications & 3 & 94.4\% & 3 & 73.3\% \\
Operations/field/grid dispatch & 7 & 91.4\% & 5 & 71.3\% \\
Compliance/policy acknowledgements \& surveys & 5 & 90.7\% & 5 & 72.7\% \\
IT maintenance / access continuity & 4 & 95.0\% & 5 & 77.3\% \\
Account/security notifications (IT/auth) & 9 & 91.9\% & 5 & 76.0\% \\
Corporate communications / executive announcements & 2 & 85.0\% & 2 & 70.0\% \\
Finance/AP/vendor billing & 8 & 86.7\% & 8 & 72.5\% \\
Travel/calendar logistics & 6 & 88.3\% & 4 & 74.2\% \\
HR/payroll/benefits administration & 6 & 86.1\% & 6 & 75.6\% \\
Employee wellness challenge & 4 & 92.5\% & 3 & 82.2\% \\
Voicemail/communications system alerts & 1 & 83.3\% & 3 & 74.4\% \\
Employee giving / donations campaign & 3 & 86.7\% & 3 & 80.0\% \\
Shipping/delivery notifications & 5 & 84.0\% & 5 & 77.3\% \\
Document sharing/collaboration prompts & 6 & 91.1\% & 3 & 86.7\% \\
Survey / prize promotion & 3 & 77.8\% & 4 & 88.3\% \\
\hline
\textbf{Total} & \textbf{77} & \textbf{88.8\%} & \textbf{69} & \textbf{74.7\%} \\
\hline
\end{tabular}%
}
\end{table}

\subsection{Confidence and Classification Accuracy}

\begin{figure}[t]
    \centering
    \includegraphics[width=0.75\linewidth]{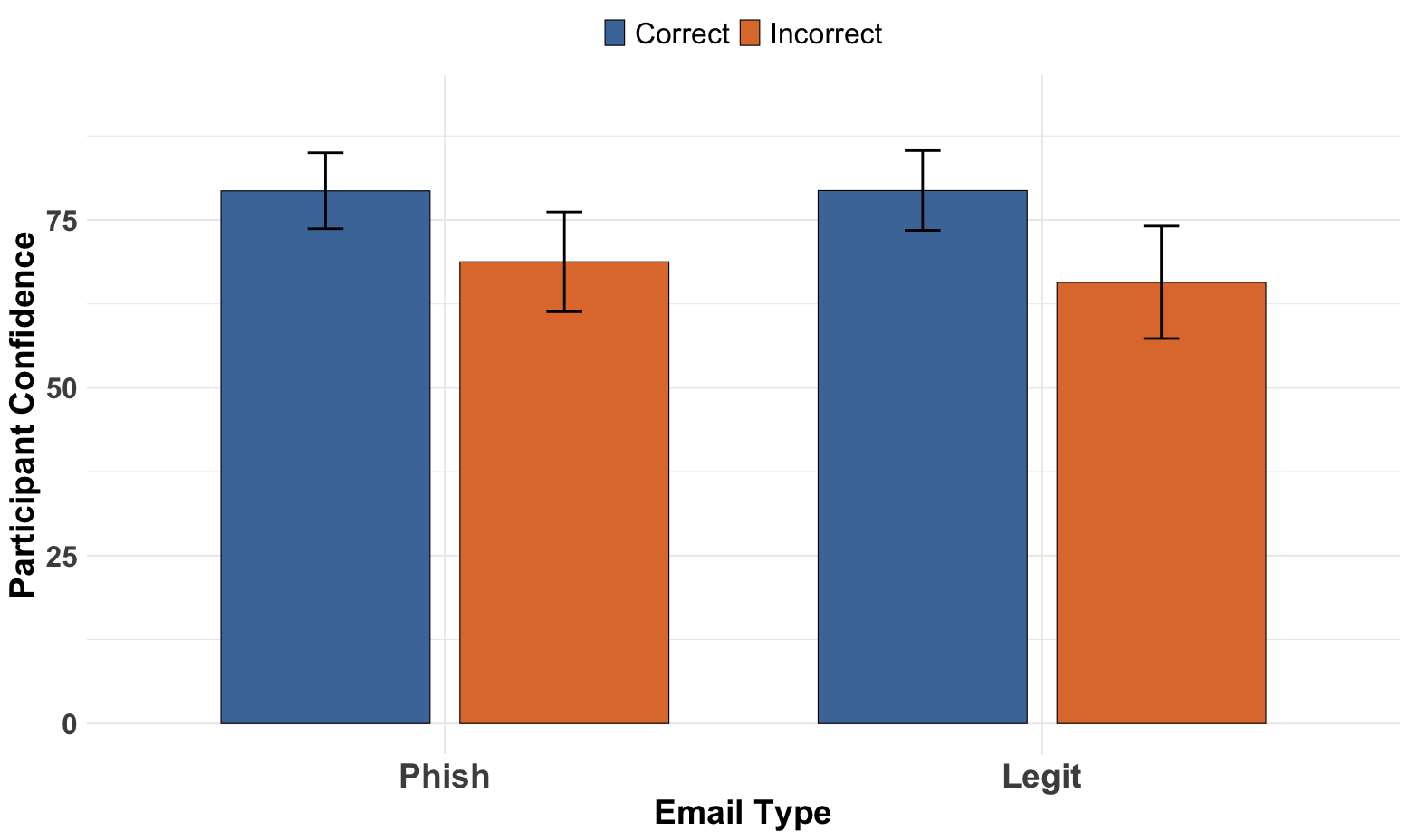}
    \caption{Mean participant confidence for phishing and legitimate emails, shown separately for correct and incorrect categorizations. Error bars represent 95\% confidence intervals across participants. Confidence was higher for correct than incorrect categorizations for both phishing and legitimate emails, whereas overall confidence did not differ substantially by true email label.}
    \label{fig:confidence_accuracy}
\end{figure}

To examine whether participants were sensitive to the reliability of their own judgments, we analyzed confidence as a function of classification accuracy. For each email, participants rated their confidence on a scale of 0-100 in increments of 10. A mixed-effects linear model predicting confidence from trial-level accuracy, with random intercepts for participant and email, showed that confidence was significantly higher for correct than incorrect responses, \(\chi^2(1)=157.03\), \(p<.001\), Figure ~\ref{fig:confidence_accuracy}. Correct responses were associated with an estimated 9.18-point increase in confidence relative to incorrect responses (\(\beta=9.18\), \(SE=0.72\)). Consistent with this pattern, mean confidence was higher for correct responses (\(M=80.1\), \(SD=22.2\)) than for incorrect responses (\(M=70.3\), \(SD=24.1\)). Confidence did not differ significantly as a function of true email label alone, \(\chi^2(1)=1.03\), \(p=.311\). In addition, the interaction between classification accuracy and true email label was not significant, \(\chi^2(1)=1.89\), \(p=.169\), indicating that the relationship between confidence and correctness was similar for phishing and legitimate emails. Taken together, these human-evaluation results suggest that the dataset is useful for studying both phishing-classification performance and the roles of cue structure, workplace context, and subjective certainty in phishing-related decision making.

\section{Classifier}
To evaluate whether the dataset is suitable for supervised phishing detection and machine learning benchmarking, we trained six models and tested their ability to distinguish phishing emails from legitimate emails. Specifically, we used Logistic Regression (LR) and Random Forest (RF) classifiers to represent both linear and non-linear classification approaches, allowing us to examine whether the dataset supports learning across different modeling strategies.

Each classifier was trained under three feature conditions: structural features only, text features only, and a combination of both. Structural features consisted of binary indicator variables capturing properties of each email (see Table ~\ref{tab:cue_distribution_final}). Text features were created by concatenating the sender, subject line, and body of each email, then converting the resulting text into numerical representations using TF-IDF vectorization. Evaluating all three feature conditions allowed us to test whether the dataset captures multiple dimensions relevant to phishing detection.

We assessed model performance using 10-fold stratified cross-validation and report the mean and standard deviation for accuracy, precision, recall, F1 score, and AUC (see Table \ref{tab:models}). To prevent data leakage, all feature extraction and vectorization steps were performed separately within each training fold.

\begin{table}[h]
\centering
\caption{Classification results (mean \(\pm\) standard deviation) across 10-fold stratified cross-validation}
\label{tab:models}
\resizebox{\textwidth}{!}{%
\begin{tabular}{llllll}
\toprule
Model & Accuracy & Precision & Recall & F1 & AUC \\
\midrule
LR: features only & 0.774 \(\pm\) 0.052 & 0.815 \(\pm\) 0.126 & 0.710 \(\pm\) 0.129 & 0.743 \(\pm\) 0.077 & 0.894 \(\pm\) 0.052 \\
LR: text only & 0.753 \(\pm\) 0.094 & 0.902 \(\pm\) 0.125 & 0.548 \(\pm\) 0.172 & 0.663 \(\pm\) 0.148 & 0.869 \(\pm\) 0.101 \\
LR: combined & 0.830 \(\pm\) 0.053 & 0.876 \(\pm\) 0.110 & 0.769 \(\pm\) 0.113 & 0.807 \(\pm\) 0.064 & 0.917 \(\pm\) 0.051 \\
RF: features only & 0.754 \(\pm\) 0.062 & 0.765 \(\pm\) 0.100 & 0.705 \(\pm\) 0.187 & 0.716 \(\pm\) 0.119 & 0.878 \(\pm\) 0.062 \\
RF: text only & 0.959 \(\pm\) 0.065 & 0.971 \(\pm\) 0.059 & 0.943 \(\pm\) 0.095 & 0.955 \(\pm\) 0.070 & 0.994 \(\pm\) 0.013 \\
RF: combined & 0.965 \(\pm\) 0.065 & 0.971 \(\pm\) 0.059 & 0.957 \(\pm\) 0.091 & 0.963 \(\pm\) 0.070 & 0.990 \(\pm\) 0.024 \\
\bottomrule
\end{tabular}%
}
\end{table}

The results indicate that the dataset is well suited for supervised phishing detection and that performance depends on both the feature representation and the classifier. Across models, combining structural and text features generally produced the strongest performance, suggesting that the two feature types capture complementary information relevant to phishing detection. This pattern was especially clear for Logistic Regression, where the combined model improved over both single-feature conditions across all major metrics. For Random Forest, the combined model also achieved the best overall performance, although the improvement over text-only features was small, indicating that much of the predictive signal may already be present in the email text.

Additionally, Random Forest outperformed Logistic Regression overall, particularly in the text-only and combined conditions. This finding suggests that the dataset contains signals that are more effectively captured by Random Forest under the modeling choices considered here. At the same time, these results should be interpreted as a proof-of-concept rather than a definitive comparison of model classes, since only two classifiers were evaluated. More broadly, the findings show that the dataset supports supervised learning and is capable of distinguishing meaningful differences in performance across model and feature configurations.

\section{Discussion}
The present study presents a synthetic phishing email dataset to support both behavioral and machine-learning research in enterprise-like contexts. The results suggest that the Nebulon dataset succeeds in several important respects. The emails span a broad set of enterprise-relevant themes, allowing the dataset to move beyond narrow consumer-oriented phishing scenarios such as generic account warnings, package-delivery scams, or prize notifications. More specifically, the stimuli reflect workplace communication contexts such as finance, compliance, information technology, travel, document sharing, physical access, and corporate announcements. Phishing plausibility often depends on whether a message fits the routines, expectations, and workflows of a particular organizational setting \citep{downs2006decision}, so broader thematic coverage can improve the general realism of the dataset.

Additionally, the dataset was designed to systematically vary phishing-related cues while preserving the overlap between phishing and legitimate messages. Because legitimate and phishing emails were often constructed as matched pairs within the same general scenario, the dataset supports comparisons that are less confounded by topic alone. Importantly, this matched-pair structure was intended to align subject matter, tone, and contextual framing across classes, but it was not designed to guarantee equal empirical difficulty for phishing and legitimate emails. The resulting emails span a range of classification difficulty, including relatively obvious phishing attempts and more ambiguous legitimate messages that share superficially suspicious features. This variation is reflected in the human-subject results: participants performed above chance overall, phishing emails were harder to classify than legitimate emails, and the signal-count analysis showed that additional phishing-related cues increased accuracy for phishing emails but decreased accuracy for legitimate emails. Taken together, these findings suggest that the dataset captures a useful range of difficulty rather than consisting only of trivially easy examples.

A further strength of the dataset is that it supports both human-judgment and machine-learning uses: participant results show that the emails function as meaningful human judgment stimuli, and classifier benchmarks show that the dataset contains learnable structure for supervised phishing detection. This helps bridge phishing research focused on human susceptibility and machine-learning evaluation. At the same time, the dataset is not intended to be unconditionally preferable to existing resources, many of which are better suited for other purposes such as large-scale modeling, naturally occurring email analysis, or webpage- and URL-focused phishing detection. Its contribution is more specific: a publicly shareable synthetic enterprise email corpus designed for matched phishing-versus-legitimate comparisons, controlled cue variation, and both human and machine evaluation.

Looking more closely at the evaluation results, both the human and classifier findings help clarify the kind of signal the dataset contains and the kinds of judgments it supports. On the human side, participants were less accurate on phishing emails than on legitimate emails, consistent with the possibility that many phishing messages resembled ordinary workplace communication rather than being immediately obvious attacks. At the same time, phishing-related signal count affected performance differently depending on the true class of the email: additional signals made phishing emails easier to detect but made legitimate emails more likely to be misclassified. This pattern suggests that the cues in the dataset were functioning less as simple rules and more as context-dependent signals whose interpretation depended on the surrounding message and scenario.

Confidence ratings and classifier performance point in a similar direction. Participants were more confident when they were correct than when they were incorrect, suggesting at least some sensitivity to the reliability of their own judgments even though classification was not perfect. To determine whether the dataset was learnable in a supervised setting, we trained two standard classifiers under multiple feature conditions and found that both performed well above chance. Performance also varied systematically across feature conditions, with combined text and structural features generally producing the strongest results, suggesting that the dataset captures complementary dimensions of the phishing detection problem. Random Forest outperformed Logistic Regression overall, although this finding should be interpreted cautiously given the limited set of models evaluated. These analyses were intended to demonstrate that the dataset supports supervised learning, not to establish an optimal modeling approach. Their purpose was to establish that the dataset is usable for supervised learning and comparative benchmarking, not to identify an optimal modeling approach. At the same time, these results should not be taken to mean that humans and models are relying on the same information in the same way. Rather, they suggest that the dataset contains multiple kinds of signal that are useful for both behavioral analysis and supervised phishing detection.

Rather than relying on unconstrained one-shot prompting, the generation process began from a previously developed synthetic dataset and was guided by subject matter expert input, enterprise-relevant themes, and an explicit taxonomy of phishing-related cues. Legitimate and phishing emails were often generated as matched pairs, which helped reduce topic-level confounds and made it possible to compare messages that were similar in scenario but different in phishing-related structure. Additional rounds of prompting, editing, and internal review were then used to reduce repetition, improve realism, and better balance cue prevalence across the dataset.

This human-in-the-loop process is important because synthetic phishing datasets can otherwise become overly template-like or rely on shallow stylistic patterns introduced during generation rather than realistic phishing signals. Instead, generated emails were repeatedly reviewed for plausibility, duplication, ambiguity, and cue coherence, and problematic examples were revised or removed before the final participant-facing set was established. The manual downselection step was especially important for removing emails whose ambiguity came more from construction artifacts than from meaningful difficulty. As a result, the final dataset reflects a compromise between realism and experimental control: it remains synthetic and structured, but it was designed to approximate workplace communication more closely than a fully automated generation pipeline would likely allow.

\subsection{Limitations}

A number of limitations should be considered when interpreting the present dataset and results. Most importantly, the emails are fully synthetic rather than naturally occurring organizational communications. Although this was necessary in order to support privacy, public release, and experimental control, synthetic generation may still introduce stylistic regularities or artifacts that differ from real enterprise email. Similarly, because the dataset was generated with LLM assistance, some messages may reflect latent model tendencies even after manual editing and review. The human-in-the-loop generation process was intended to reduce these issues, but it cannot eliminate them entirely.

The human evaluation context was also limited in important ways. Emails were judged individually in an online task rather than within a realistic inbox, workflow, or organizational environment, so participants did not have access to many contextual cues that influence real-world phishing decisions. In addition, the participant sample was modest, and some enterprise themes and cue combinations were represented by relatively few items. The cue taxonomy itself was intentionally compact rather than exhaustive, which improved interpretability but means that not all potentially relevant phishing signals were explicitly annotated. Finally, the classifier analyses should be interpreted as proof-of-concept benchmarks rather than a comprehensive assessment of phishing detection methods. Only two model families and a limited set of feature representations were evaluated, so the present results establish that the dataset is usable for supervised learning but do not identify an optimal modeling approach or demonstrate generalization to naturally occurring enterprise emails. Nonetheless, the dataset fills an important gap by providing a controlled, publicly shareable set of enterprise-oriented phishing and legitimate emails that can support both behavioral and machine-learning research.

\section{Public Release}
The dataset is publicly released through Kaggle to support reuse, benchmarking, and extension by other researchers:
\url{https://www.kaggle.com/datasets/emilywinokur/synthetic-email-corpus}.

Although participants evaluated the broader 169-email candidate set during the study, the public release and all reported analyses are based on the final retained set of 146 emails. The release includes five files:

\begin{itemize}
    \item \texttt{seed\_dataset.csv}, which contains the initial synthetic seed dataset used as a structured starting point during generation;
    \item \texttt{final\_email\_dataset.csv}, which contains the final 146-email retained email-level dataset used for analysis and public release;
    \item \texttt{email\_level\_human\_summary.csv}, which provides email-level aggregate participant response measures for the retained 146-email set;
    \item \texttt{participant\_responses\_anonymized.csv}, which contains de-identified participant-level response data from the human evaluation study for the retained 146-email set; and
    \item \texttt{data\_dictionary.xlsx}, which documents the variables included in each released file.
\end{itemize}

Together, these files are intended to support several types of reuse. The final email dataset provides the main corpus for phishing detection and benchmarking, the participant-response file supports behavioral analyses, the email-level summary file supports item-level difficulty and confidence analyses, and the seed dataset improves transparency around the structured starting point used during generation. Because the release was designed for public sharing, all emails are synthetic and tied to the fictitious Nebulon organization rather than any real company or operational environment, and participant identifiers have been anonymized prior to release.

\section{Conclusion}
In this work, we introduced a publicly shareable synthetic enterprise phishing email dataset designed to support both behavioral phishing research and machine-learning benchmarking. The dataset combines realistic workplace themes, matched legitimate and phishing scenarios, controlled cue variation, and validation in both human-subject and classifier settings. Although it is not a substitute for naturally occurring organizational email, it provides a useful and controlled resource for studying phishing detection, human susceptibility, and benchmark performance in enterprise-like contexts.

\section{Funding}

\textit{This article has been authored by employees of National Technology \& Engineering Solutions of Sandia, LLC under Contract No. DE-NA0003525 with the U.S. Department of Energy (DOE). The employee owns all right, title and interest in and to the article and is solely responsible for its contents. The United States Government retains and the publisher, by accepting the article for publication, acknowledges that the United States Government retains a non-exclusive, paid-up, irrevocable, world-wide license to publish or reproduce the published form of this article or allow others to do so, for United States Government purposes. The DOE will provide public access to these results of federally sponsored research in accordance with the DOE Public Access Plan \href{https://www.energy.gov/downloads/doe-public-access-plan}{https://www.energy.gov/downloads/doe-public-access-plan}.}

\bibliographystyle{plain}
\bibliography{references}

\newpage
\appendix
\section*{Appendix}
\appendix
\renewcommand{\thefigure}{A.\arabic{figure}}
\renewcommand{\thetable}{A.\arabic{table}}
\setcounter{figure}{0}
\setcounter{table}{0}

\section{Training Modules}
\label{apx:training_modules}

\begin{figure}[H]
    \centering

    \begin{subfigure}[b]{\textwidth}
        \centering
        \includegraphics[width=\textwidth,height=0.35\textheight,keepaspectratio]{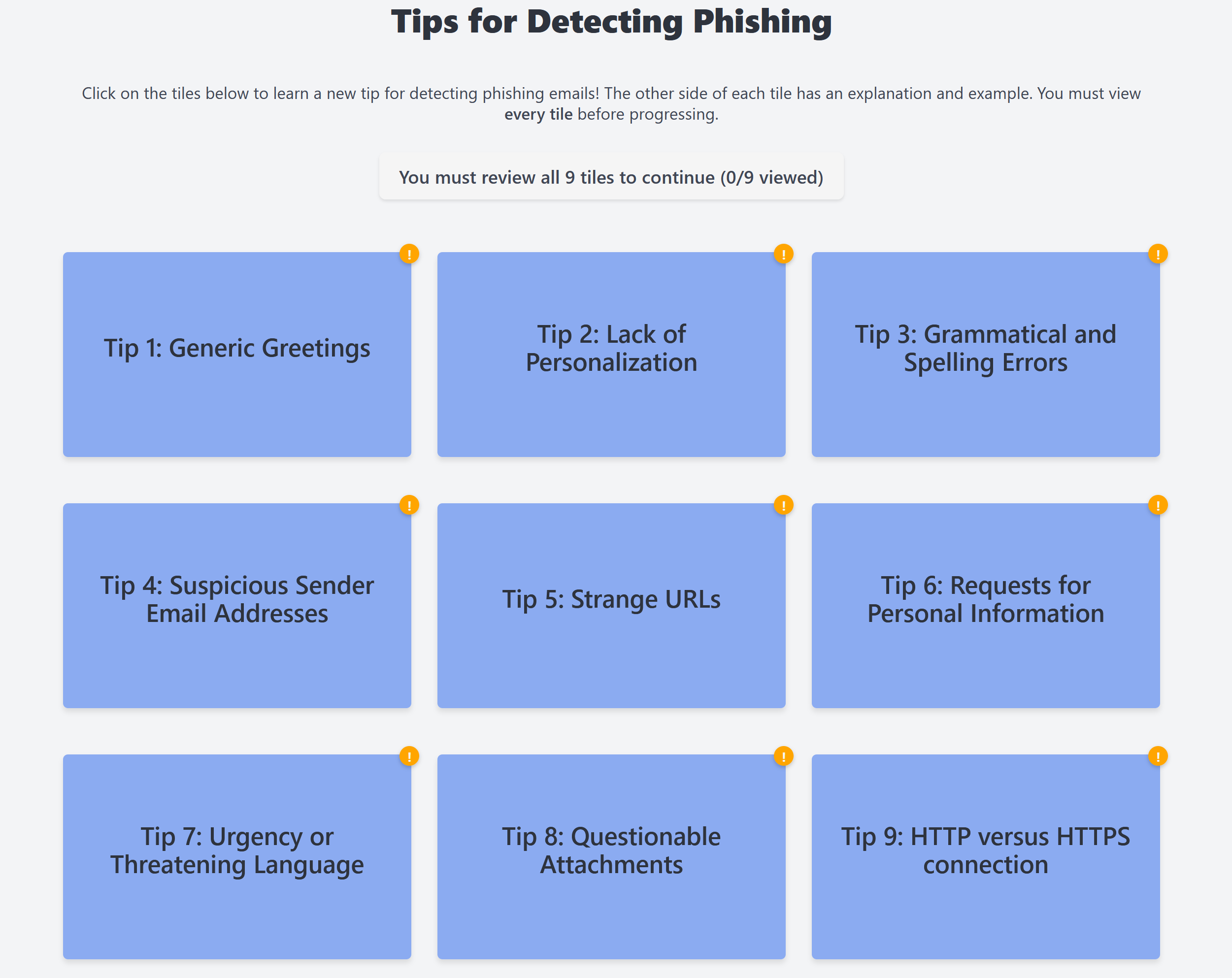}
        \caption{Training module interface.}
        \label{fig:training_module_a}
    \end{subfigure}

    \vspace{1em}

    \begin{subfigure}[b]{\textwidth}
        \centering
        \includegraphics[width=\textwidth,height=0.35\textheight,keepaspectratio]{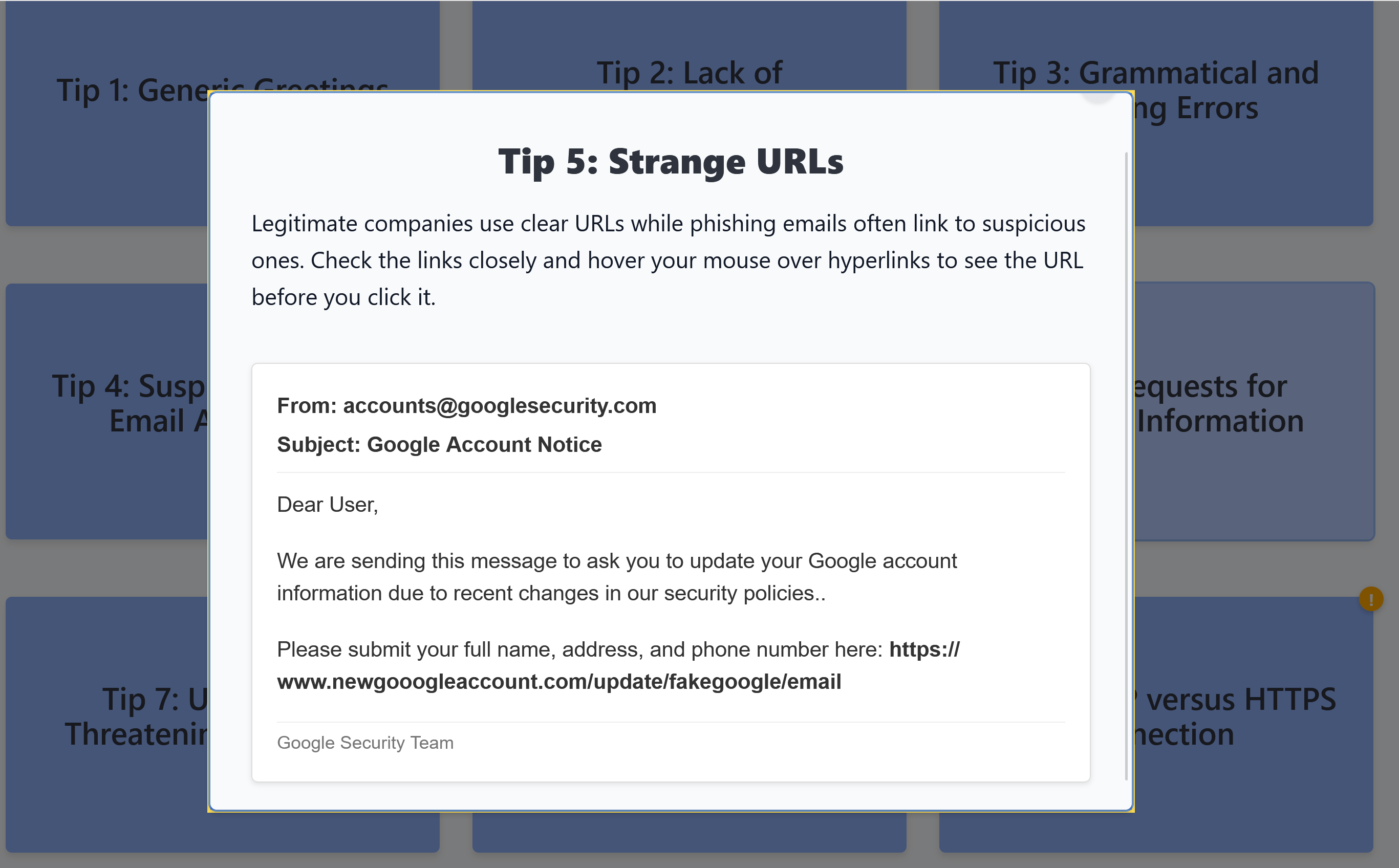}
        \caption{Example training screen.}
        \label{fig:training_module_b}
    \end{subfigure}

    \caption{Training module used in the experiment. (a) The interface presented nine interactive tips to help participants identify phishing emails. (b) Clicking a tile displayed information about the selected tip. Participants were required to click on all tiles before proceeding to the experiment.}
    \label{fig:training_module}
\end{figure}

\subsection{Training Comprehension Quiz}
\label{apx:trainingquiz}

\begin{enumerate}
    \item \textbf{What type of greeting is commonly found in phishing emails?}
    \begin{enumerate}[label=\Alph*.]
        \item Greetings that include your account information
        \item Personalized greetings that include your name
        \item Generic greetings such as ``Dear customer''
        \item No greeting at all
        \item Both C and D
    \end{enumerate}
    \textbf{Correct Answer:} E) Both C and D

    \item \textbf{Which of the following are more likely to be present in a phishing email than in a legitimate email? (Select all that apply)}
    \begin{enumerate}[label=\Alph*.]
        \item Grammatical and/or spelling errors
        \item A sender email address that resembles but is not the same as a legitimate company
        \item A clear and straightforward URL
        \item A request for your Social Security number
        \item A lack of urgency
    \end{enumerate}
    \textbf{Correct Answers:} A) Grammatical and/or spelling errors; B) A sender email address that resembles but is not the same as a legitimate company; D) A request for your Social Security number

    \item \textbf{What type of language is often used in phishing emails to provoke immediate action?}
    \begin{enumerate}[label=\Alph*.]
        \item Polite and courteous language
        \item Technical jargon
        \item Urgent or threatening language
        \item Formal business language
    \end{enumerate}
    \textbf{Correct Answer:} C) Urgent or threatening language

    \item \textbf{Which of the following statements is true regarding requests for personal information in emails?}
    \begin{enumerate}[label=\Alph*.]
        \item Legitimate companies often ask for sensitive information through email.
        \item Phishing emails may ask you to verify your account by providing personal details.
        \item You should always provide your personal information if the email looks official.
        \item It is safe to respond to emails requesting personal information if you recognize the sender.
    \end{enumerate}
    \textbf{Correct Answer:} B) Phishing emails may ask you to verify your account by providing personal details.

    \item \textbf{True or False: It is safe to click on suspicious links in emails if the email looks legitimate.}
    \begin{enumerate}[label=\Alph*.]
        \item True
        \item False
    \end{enumerate}
    \textbf{Correct Answer:} B) False

    \item \textbf{Which of the following statements is TRUE regarding suspicious email attachments?}
    \begin{enumerate}[label=\Alph*.]
        \item Attachments with common file types like .docx and .pdf are always safe to open.
        \item An attachment with a double extension, such as ``document.pdf.exe,'' is likely to be malicious.
        \item Emails from known contacts can never contain malicious attachments.
        \item It is safe to open attachments if the email contains a friendly message.
    \end{enumerate}
    \textbf{Correct Answer:} B) An attachment with a double extension, such as ``document.pdf.exe,'' is likely to be malicious.

    \item \textbf{Why is it important to prefer URLs that use HTTPS over HTTP when accessing websites?}
    \begin{enumerate}[label=\alph*.]
        \item Because HTTP websites load faster than HTTPS websites
        \item Because HTTP transmits data in plain text, making it vulnerable to interception, while HTTPS encrypts data for security
        \item Because HTTPS websites do not require passwords
        \item Because HTTP websites are always unsafe and never legitimate
    \end{enumerate}
    \textbf{Correct Answer:} B) HTTP sends data in plain text, allowing attackers to intercept sensitive information. HTTPS encrypts data during transmission, making it much harder for attackers to access or tamper with the information.
\end{enumerate}

\subsection{Signal Count Distribution Across Legitimate and Phishing Emails}

\begin{table}[H]
\centering
\caption{Distribution of retained emails by phishing-related signal count and true email label. Percentages are reported within each email label.}
\label{tab:supp_signal_count_distribution}
\renewcommand{\arraystretch}{1.2}
\begin{tabular}{rcccc}
\hline
\textbf{Signal count} & \textbf{Legit \(n\)} & \textbf{Legit \%} & \textbf{Phish \(n\)} & \textbf{Phish \%} \\
\hline
0 & 6  & 7.8\%  & 0  & 0.0\% \\
1 & 30 & 39.0\% & 2  & 2.9\% \\
2 & 33 & 42.9\% & 18 & 26.1\% \\
3 & 7  & 9.1\%  & 18 & 26.1\% \\
4 & 1  & 1.3\%  & 25 & 36.2\% \\
5 & 0  & 0.0\%  & 5  & 7.3\% \\
6 & 0  & 0.0\%  & 1  & 1.5\% \\
\hline
\end{tabular}
\end{table}

\subsection{Accuracy by Theme and True Label}

\begin{figure}[H]
    \centering
    \includegraphics[width=0.95\linewidth]{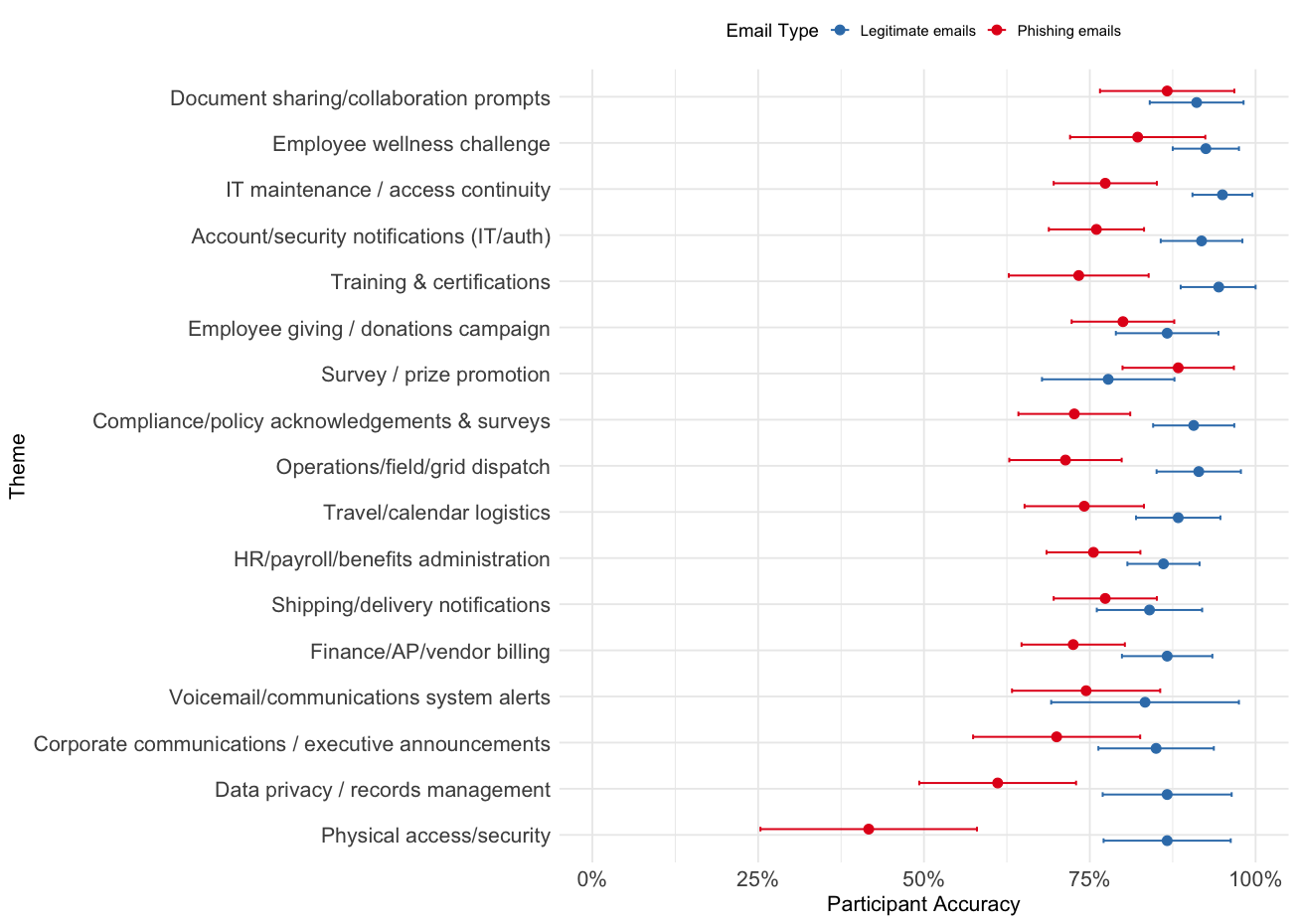}
    \caption{Mean participant accuracy by enterprise theme and true email label. Points show mean participant accuracy within each theme for legitimate and phishing emails, and horizontal error bars represent 95\% confidence intervals across participants. In most themes, legitimate emails were classified more accurately than phishing emails, although the magnitude of this difference varied across workplace contexts.}
    \label{fig:theme_accuracy_by_label_supp}
\end{figure}

\end{document}